\documentstyle[prl,aps]{revtex}
\begin{document}
\title{Compacton-like  Solutions for Modified KdV and other 
Nonlinear Equations } 
\author{C.~Nagaraja Kumar and Prasanta K.~Panigrahi, \\ School
of Physics,\\ University of Hyderabad,\\ Hyderabad 500 046,
India.\\ }
\maketitle
\def\be{\begin{equation}}
\def\ee{\end{equation}}
\def\kmn{$K(m,n)$}
\vskip 0.5truecm
\begin{abstract}
 We present compacton-like solution of the modified KdV equation
and compare its properties with those of the compactons and
solitons. We further show that, the  nonlinear Schr{\"o}dinger
equation with a source term and  other higher order KdV-like
equations also possess compact solutions of the similar form.

\end{abstract}
\draft
 \newpage
Compactons are new class of localized solutions for families of
fully nonlinear, dispersive, partial differential equations.
Unlike the solitons, which, although highly localized, still
have infinite span, these solutions have compact support; they
vanish identically outside a finite  region. Hence, these
solitary waves have been christened as  compactons \cite{RH93}.
Remarkably, there is  strong numerical evidence that, the
collision of two compactons is elastic, a feature characterizing
the solitons. These equations, arising in the context of pattern
formation in nonlinear media,  seem to have only a finite number
of local conservation laws; yet, the behaviour of the solutions
closely mimic those of the solitons of the integrable models.

The first, two parameter family of fully nonlinear, dispersive,
equations \kmn,    admitting compacton solutions, are of the
form,
 
\begin{equation}
u_t +(u^m)_x +(u^n)_{3x} =0 \qquad\qquad m > 0, \qquad\qquad 1 < n < 3 , 
\label{roseq}
\end{equation}
with $u_t \equiv \frac{\partial u }{\partial t} $ and $u_x \equiv \frac
{\partial u }{\partial x} $.  These
equations arose in the process of understanding the role of
nonlinear dispersion, in the formation of structures like liquid
drops  \cite{OR89}. The compacton solution of $K(2,2) $ reads,
\be
 u_c  =  { {4\lambda}\over 3} \cos^2( {{x -\lambda t }\over
4})\quad , \label{rossol}
\ee
when $\mid x - \lambda t \mid \le 2 \pi $ and $u_c = 0 $,
otherwise.\\

Unlike the solitons, the width here is independent of velocity;
however, the amplitude  depends on it. It has been shown that,
$K(2,2)$ admits only four  local conservation laws.  Some of the
other representative compactons are,
\be
u_c = [ 37.5 \lambda - ({x - \lambda t })^2]/30 \quad,
\ee
and
\be
u_c = \pm \sqrt{[3 \lambda / 2 ]} cos (({x - \lambda t })/3)  \quad;
\ee
these are the solutions  of $K(3,2)$, and $ K(3,3)$ equations,
respectively.

The  $K ( m,n)$  family is not derivable from a first order
Lagrangian, except for $ n = 1$ \cite{CSS}.  A  generalized
sequence of KdV-like equations, which could be given  a
Lagrangian formulation, have also been shown to admit compacton
solutions.  These equations
\be
u_t +  u_x u^{l-2} + \alpha [ 2 u_{3x} u^p + 4 p u^{p-1} u_x
u_{2x} + p ( p - 1 ) u^{p-2} (u_x)^3 ] = 0 \quad, \label{csseq}
\ee
\noindent
have the same terms as  in Eq.(\ref{roseq}); the relative
weights of the terms are different.  Further generalizations to,
one parameter generalized KdV equation \cite{KC}, two parameter
odd order KdV equations \cite{BD98}, enlarged the class of
evolution equations, which admitted  solutions with compact
support.  These type of solutions have also appeared in the
context of baby Skyrmions \cite{GP}.

The stability of the compacton solutions was considered in Ref.
\cite{BK98}; it was shown,  by linear stability analysis, as
well as, by Lyapunov stability criterion, that, these solutions
are stable for arbitrary values of the nonlinear parameters.

None of the  evolution equations possessing compacton solutions
have been shown to be integrable; infact, some are
non-integrable and possess only finite number of conserved
quantities. Hence, it is of great interest to search for
compacton-like solutions of the integrable nonlinear equations.
In particular, one would like to compare their properties with
the solitons on one hand, and the compactons on the other.
Furthermore, the possibility of these compact solutions arising
from  the nonlinear equations, relevant for physical problems,
will make them amenable for experimental detection.

In this note, we first show the existence of compacton-like
solution to the modified KdV  (MKdV) equation \cite{Das},

\be
 u_t + u^2 u_x + u_{3x} = 0 \quad.\label{mkdv}
\ee
 The solution is
of the form,

\def\amp{\frac{\sqrt{32}}{3} k}  

\be
u_c (x , t ) = \amp \frac{\cos^2 k ( x - 4 k^2 t)}{ (1 - \frac{2}{3}
\cos^2 k ( x - 4 k^2 t))} \quad,
\label{comp}
\ee
in the region $ \mid ( x - 4 k^2 t) \mid \le \frac{\pi}{2k} $
and zero otherwise.\\
\noindent
Note that, for this compact solution, both  the width and amplitude
are proportional to the square root of the velocity. This behaviour
is reminiscent of the solitons.
Though the second derivative of the solution is discontinuous
at the boundaries, it is a strong solution of the equation of
motion, a feature
similar to the compactons.  We would like to  point
out that, $u_c  ( x, t)$, when unconstrained is a periodic solution
of the  MKdV equation, which 
to the best of our knowledge, has not appeared in the  
literature earlier.  Judicious truncation of the periodic solution, by
confining it to the fundamental strip, gives the compact support
for this solution.  Though this is a solution of the integrable
equation, one needs to be careful with the conserved quantities, in light
of the fact that some derivatives of the solution are
discontinuous at the boundaries.

The Hamiltonian from which MKdV equation  can be derived
by variational principle,
\be 
H = \frac{1}{2} \int_{-\infty}^\infty u_x^2 d x -
\frac{1}{12} \int_{-\infty}^\infty u^4 d x \quad , \label{ham}
\ee
and the momentum expression,  given by, 
\be
P = \frac{1}{2} \int_{-\infty}^\infty u^2 d x \quad , \label{mom}
\ee
are well defined at the edges.

Explicitly, for the above solution, the energy and momentum are
given, respectively by, $E_c = -~(16/3)~ k^3 $ and $ P_c
= 4~ \pi~ k $. For the soliton solution of the MKdV equation
\be
u_s (x,t)  = \sqrt{6}~k~ sech k ( x - k^2 t ) \quad,\label{mkdvsol} 
\ee
the corresponding quantities are 
$E_s =   - 2~ k^3 $ and $ P_s =   6~ k$.
Hence, the compacton for postive values of $k$, has lower 
energy compared to the soliton.  The situation is opposite for negative
values of $k$. 
Other conserved quantities,  involving the even derivatives of the 
solutions, are not well defined at the boundaries. In this sense,
these solutions are similar to compactons.

Next, we show that this compacton-like  solution of the MKdV
equation is also a strong solution of the  nonlinear
Schr{\"o}dinger equation (NLSE) and higher order KdV-like
equations. 

The solution with compact support obeys the NLSE with a source
term:
\be
i q_t + \frac{1}{2} q_{xx} + \mid q \mid^2 q -  \eta = 0\quad . \label{nls}
\ee
Using the following anstaz, 
\be
q ( x , t ) = e^{i[\psi(\xi) - \omega t ]} a(\xi) \quad , \label{anst} 
\ee
where $\xi = x - v t $, and choosing the source term as $\eta
(\xi) = K  e^{i[\psi(\xi) - \omega t ]}$, we can  separate the
real and the imaginary parts of the equation as,
\be
v \psi^\prime a + \omega a + \frac{a^{\prime\prime}}{2} -
\frac{{{\psi^\prime}^2} a }{2} + a^3 - K = 0 \quad,
 \label{real}
\ee  
\be
- v a^\prime +  \frac{\psi^{\prime\prime} a}{2}  + \psi^\prime a^\prime
= 0 \quad .
\label{imag}
\ee
Eq.(\ref{imag}) can be straightforwardly solved to give
\be
\psi^\prime = v + \frac{P}{a^2} \quad , \label{sol}
\ee
where $P$ is the integration constant. Choosing $P = 0 $, we
 arrive at
the following solutions for the functions $\psi(\xi)$ and $a(\xi)$:
\be
\psi(\xi)  = v \xi \quad ,
\ee
and
\be
a(\xi) = (\frac{16 K}{27})^{1/3} \frac{\cos^2 [(\frac{27}{16})^\frac{1}{6}
K^\frac{1}{3}
(x - v t)]}{ (1 - \frac{2}{3} \cos^2  [(\frac{27}{16})^\frac{1}{6}
K^\frac{1}{3} (x - v t )])} \quad ,\ee

where $\omega $ is related to $v$ by
\be
2 \omega + v^2 = -\frac{27}{4}  (16 K/27)^{2/3 }.
\ee
Note that, the solution exists for values of $\omega \le -
\frac{27}{4}  (16 K/27)^{2/3 }$.  We would like to add that, we
have taken the simplest solution of the equation involving the
phase. In principle, one can choose $ P \ne 0 $, which may give
rise to shock-wave type solutions\cite{cai}.

It is worth emphasizing that the NLSE plays a significant role
in nonlinear optics \cite{HT},\cite{GPA}.  Since $q$ there,
represents the electric field, the corresponding source term
$`\eta'$ can be understood as dipole sources. Hence, a fluid
nonlinear medium with moving dipoles can be a plausible source
of these type of solitary waves.

We  now  show that this compacton-like solution 
also satisfies  higher order KdV like equations:  
\be
u_t + ( l u + m u^4) u_x + 5 u^2 u_{3x} + p u_{5x}  = 0\quad,
\ee
where $ l= \frac{10}{3 p},\quad m= \frac{5 }{ 3 p} $
and $ k^2 = \frac{1}{4 p}$.
The solution of this equation is of the form 
\be
u_c (x , t ) = \frac{4}{3} \frac{\cos^2 k ( x -   4 k^2 t)}{ (1 - \frac{2}{3}
\cos^2 k ( x - 4 k^2 t))} \quad.
\label{compgkdv}
\ee
Note that, for this solution the amplitude is independent of the velocity
where as the width depends on it. 
Interestingly, the following fifth order KdV-like equation, 
\be
u_t + l  u^4 u_x +  u_x^3 +  u^2 u_{3x} + q u_{5x} = 0 \quad,
\ee
has a compact solution  of the form
\be
u_c (x , t ) = \frac{2 \sqrt{10}}{3}
 \frac{\cos^2 k ( x - 4 k^2 t)}{ (1 - \frac{2}{3}
\cos^2 k ( x - 4 k^2 t))} \quad,
\label{compg2kdv}
\ee
for $ l = \frac{243}{10 Q}$ and $k^2 = \frac{1}{ 4 Q}$. 

In conclusion, we have shown the existence of solutions with
compact support for MKdV, nonlinear Schr\"odinger equation and
higher order KdV-like equations.  Since the MKdV equation
manifests in diverse physical phenomena, it will be exciting, if
these waves can be realized in an experimental situation. For
this purpose, one needs to check the stability of these
solutions. This question is currently under study. Similarly, one
can enquire about the possibilities of  these type of solutions
occurring in other NLSE type equations arising in the context of
nonlinear optics.

{\bf ACKNOWLEDGEMENTS}

The work of CNK  is supported by CSIR, INDIA, through S.R.A. Scheme.
We acknowledge  useful discussions with Prof. V. Srinivasan.

\end{document}